\documentclass[twoside,12pt]{article}
\usepackage{epsfig}

\begin{document}

\title{ \vspace{1cm} On the QCD phase structure from effective models}
\author{Bernd-Jochen Schaefer$^1$ and Mathias Wagner$^2$\\
  \\
  $^1$Institut f\"ur Physik, Karl-Franzens-Universit\"at Graz,
  Austria\\
  $^2$Institut f\"ur Kernphysik, Technische Universit\"at Darmstadt,
  Germany
}
\maketitle
\begin{abstract} 
  Some recent theoretical developments of the QCD phase diagram are
  summarized. Chiral symmetry restoration and the
  confinement/deconfinement transition at nonzero temperature and
  quark densities are analyzed in the framework of an effective linear
  sigma model with three light quark flavors. The sensitivity of the
  chiral transition as well as the existence of a critical end point
  in the phase diagram on the value of the sigma mass is explored. The
  influence of the axial anomaly on the chiral critical surface is
  addressed. Finally, the modifications by the inclusion of the
  Polyakov loop on the phase structure are investigated.
\end{abstract}
\section{Introduction} 

In recent years the phase diagram of strongly interacting matter has
become the focus of theoretical and experimental attention. It is
believed that the underlying fundamental theory of the strong
interaction is described by Quantum Chromodynamics (QCD). At finite
temperature and finite densities QCD predicts two different phase
transitions which are associated with two opposite quark mass limits
(for reviews see e.g.~\cite{MeyerOrtmanns:1996ea, Rischke:2003mt}).

For vanishing quark masses, i.e., in the chiral limit, QCD has an
exact global $U(N_f)_L \times U(N_f)_R$ chiral symmetry, where $N_f$
denotes the number of quark flavors. The axial $U(1)_A$ anomaly,
induced by instantons, breaks the axial $U(N_f)_A$ part of the chiral
symmetry explicitly to $SU(N_f)_A$. In addition, in the vacuum the
$SU(N_f)_A$ is spontaneously broken by a finite expectation value of
the quark condensate $\langle \bar q q \rangle \neq 0$. As a
consequence of the Goldstone theorem, $N_f^2-1$ massless pseudoscalar
Goldstone bosons are expected to emerge. For $N_f=2$ the associated
Goldstone bosons are the three pseudoscalar pions and for $N_f=3$ one
has additionally the four kaons and the pseudoscalar $\eta$ meson
which together with the pions constitute the pseudoscalar meson octet.
Once the quark masses obtain finite values, i.e., leaving the chiral
limit, chiral symmetry is broken explicitly and all these Goldstone
bosons acquire masses as measurable in the experiment.
However, at high temperatures and densities this symmetry breaking
pattern changes drastically. In hot and dense matter the $SU(N_f)_A$
symmetry and additionally, if instantons are sufficiently screened,
the explicitly broken axial $U(1)_A$ symmetry will become restored
again. As a consequence, the masses of the pseudoscalar Goldstone
bosons will degenerate with the masses of the corresponding chiral
scalar partners, signaling in this way the restoration of chiral
symmetry. The associated phase transition is commonly referred to as
the chiral phase transition.

In the opposite quark mass limit, the so-called quenched limit of QCD
with infinitely heavy quark masses, QCD reduces to a pure $SU(N_c)$
gauge theory which is invariant under a global $Z(N_c)$ center
symmetry. In contrast to the chiral symmetry, the center symmetry is
spontaneously broken at high temperatures and densities, i.e., in the
color deconfined quark-gluon plasma phase and is restored in the
hadronic phase at small temperatures and densities. The associated
phase transition from the hadronic (glueball) phase to the color
deconfined plasma phase is the confinement/deconfinement phase
transition. The center symmetry is always broken explicitly when
dynamical quarks are present, i.e., when the quenched limit of QCD is
left.

\begin{figure}[tb]
\begin{minipage}[t]{0.5\textwidth}
\epsfig{file=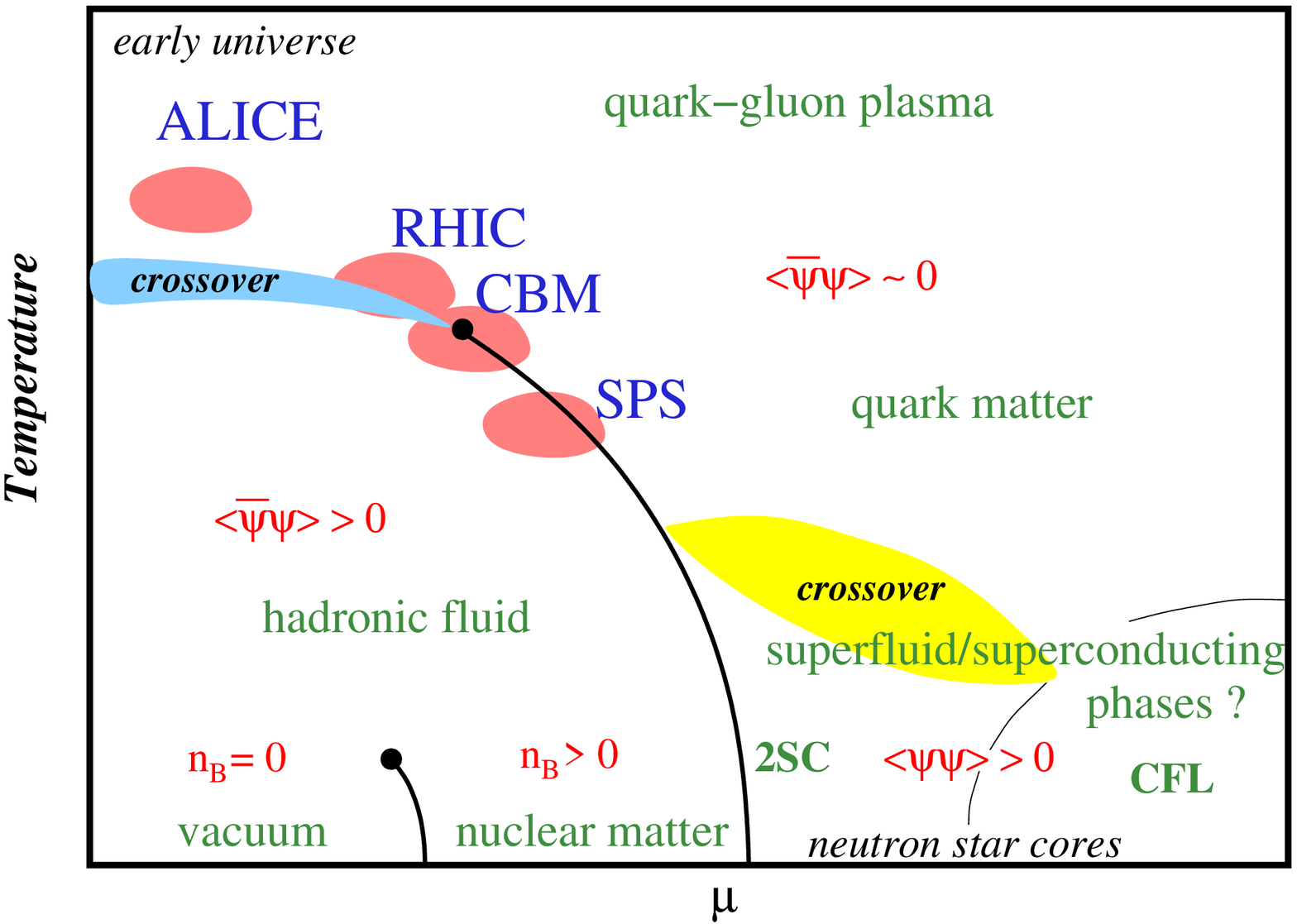,height=6.5cm}
\end{minipage}
\hspace*{1cm}
\begin{minipage}[t]{0.4\textwidth}
\epsfig{file=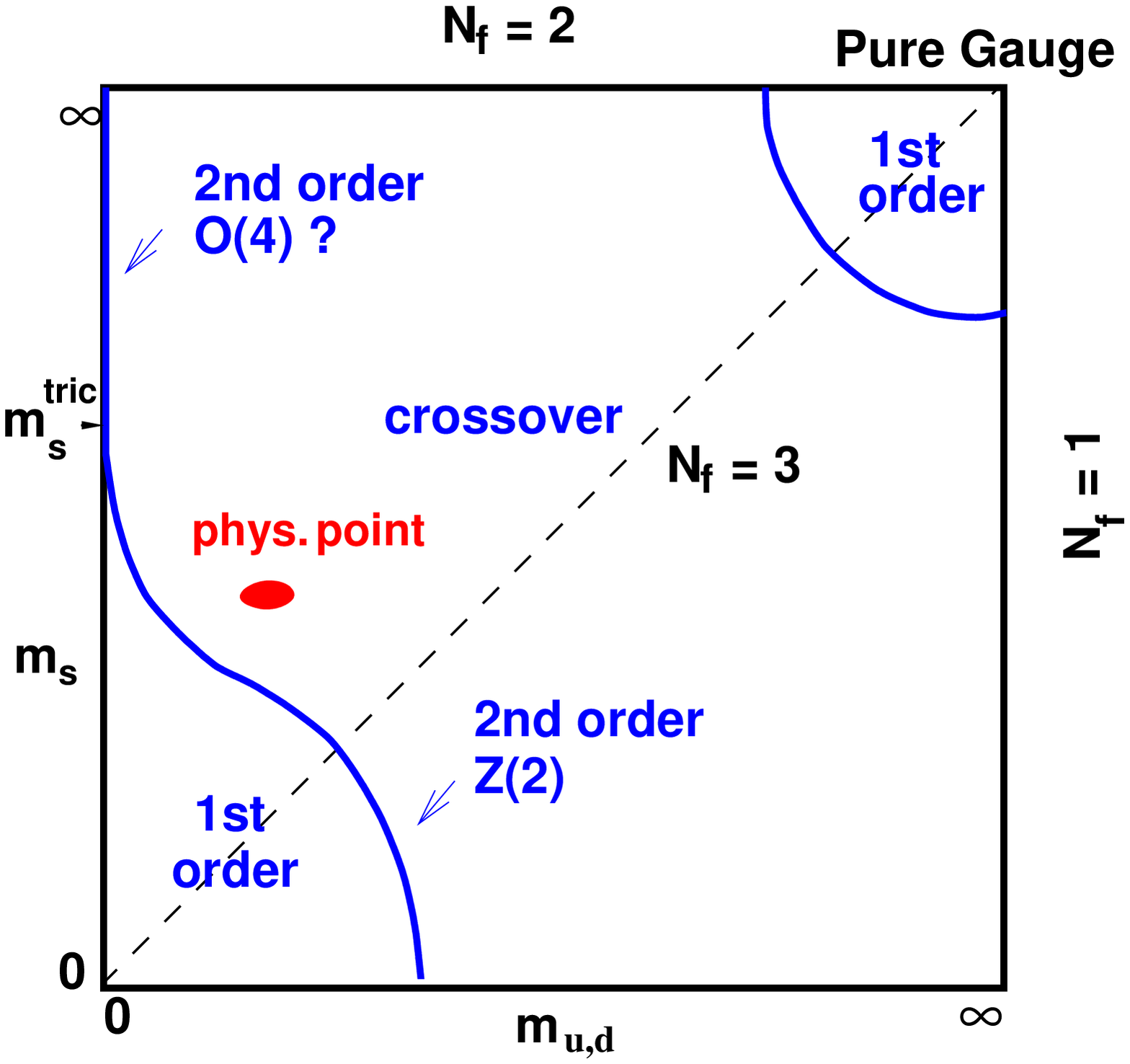,height=7cm}
\end{minipage}
\begin{center} 
\begin{minipage}[t]{\textwidth}
  \caption{Schematic phase transition behavior of $N_f = 2+1$ flavor
    QCD in the ($T,\mu$) plane (left panel) and for vanishing chemical
    potential in the ($m_{u,d}, m_s$) quark mass plane (right
    panel)\cite{Brown:1988qe}.
    \label{fig:columbia}}
\end{minipage}
\end{center}
\end{figure}

Both phase transitions are conceptually distinct phenomena of QCD. For
the experiment it is important to investigate and understand the
interplay between these phase transitions, in particular, for
realistic quark masses. Based on theoretical models and QCD lattice
simulations a generic phase diagram for $N_f = 2+1$ quark flavors in
the temperature and (baryo)chemical potential $\mu$ plane can be drawn
as in the left panel of Fig.~\ref{fig:columbia}. Here not only the
chiral and deconfinement phase transition from the hadronic fluid
phase to the quark-gluon plasma are shown. The regions probed by some
already running or planned relativistic heavy-ion collision
experiments such as ALICE, RHIC, CBM and SPS are also marked in the
figure. So far, it is still an open issue whether both phase
transitions, the chiral and deconfinement transition, take place at
the same temperatures and densities yielding thus a single transition
or crossover line in the QCD phase diagram as indicated in the figure.
For example, McLerran and Pisarksi suggested that this is not the case
at moderate temperature and large chemical potential. Based on
large-$N_c$ arguments, they concluded that in this phase diagram
region there might be a new, so-called quarkyonic phase which is still
confining but chirally symmetric \cite{McLerran:2007qj}.

The situation is even more sophisticated since some properties of the
chiral phase transition such as its order depend on $N_f$ and the
strength of the axial anomaly. The status for vanishing chemical
potential is partly summarized in the right panel of
Fig.~\ref{fig:columbia}: in the limiting case of two massless light
quarks, $m_{u,d}=0$, and an infinite strange quark mass $m_s$, which
corresponds to $N_f=2$, it is conjectured that the finite temperature
 chiral phase transition is of second-order for a
constant anomaly strength and lies in the universality class of the
Heisenberg $O(4)$ model in three dimensions.

If the anomaly strength is identified with the instanton density, a
temperature-dependent strength of the axial anomaly would arise. It is
supposed that the strength vanishes at high temperatures. This
temperature-dependent axial anomaly can change the chiral transition
to first-order. 

The chiral transition will also be of first-order once the strange
quark mass drops below a certain critical value. This critical mass
value is a tricritical end point of the second-order transition line.
For three vanishing quark masses the first-order transition has been
confirmed by a renormalization group analysis and is independent of
the strength of the axial anomaly~\cite{Pisarski:1983ms}. The
first-order region in the ($m_{u,d},m_s$) plane still persists for
small light and strange quark masses and is finally terminated by a
second-order transition boundary. This boundary line separates the
first-order region from the crossover region in the ($m_{u,d},m_s$)
plane. QCD lattice simulations indicate that the physical mass point,
labeled as a red point in the Fig.~\ref{fig:columbia}, is located in
the crossover regime. In contrast to the boundary line at $m_{u,d}=0$,
which lies presumably in the $O(4)$ universality class, the
universality class changes for finite $m_{u,d}$ to the one of the
$Z(2)$ three-dimensional Ising model.

For finite chemical potential the area of the first-order regime in
the quark mass plane also changes. If it grows for increasing chemical
potential the boundary may hit and pass over the position of the
physical mass point, turning the chiral phase transition from a
crossover to a second-order or first-order transition and as a
consequence, the existence of critical end point (CEP) in the QCD
phase diagram becomes possible. This scenario is denoted as the
standard one.

On the other hand, if the first-order region shrinks for increasing
chemical potential the chiral phase transition sticks to be a
crossover and the existence of a CEP in the phase diagram is not
possible, which is commonly labeled as the non-standard scenario and
has been predicted by de Forcrand and Philipsen \cite{Forcrand2002}.

The existence or exclusion of a QCD critical end point has not yet
been confirmed by QCD lattice simulations. At finite chemical
potential these Monte-Carlo simulations suffer from the notorious sign
problem because the quark determinant in the QCD partition function
becomes a complex quantity which entangles its probability
interpretation. But recently, some progress could be achieved in
extrapolating zero chemical potential Monte-Carlo simulations to
finite chemical potentials \cite{Schmidt:2006us}. However, all these
extrapolation techniques are still limited to small chemical
potentials. Furthermore, both above mentioned scenarios, the standard
and the non-standard ones, are seen on the lattice with different
extrapolation techniques.

So far, only model studies give direct and indirect evidences for the
existence of the critical end point in the whole phase diagram. But
they cannot predict its precise location. The location and even its
existence depend also on the magnitude of the axial anomaly and on
vector-channel interactions \cite{Struber:2007bm, Fukushima:2008is}.
This is plausible because the zeroth-component of the vector-channel
interaction is directly coupled to the quark density which surely
modifies the interactions in the finite-density environment. Due to
the repulsive vector-vector interaction the first-order transition is
weakened. As a consequence a standard scenario could change to a
non-standard one depending, of course, on the strength of the vector
coupling.

In all, the influence of the axial anomaly and also of the
vector-vector interaction on the QCD phase structure is of relevance
and should be investigated in a quantitative manner. Moreover, in
order to bridge the gap between existing lattice data at zero chemical
potential and interesting regions in the QCD phase diagram at finite
chemical potential effective models are useful and inevitable tools.
They share the relevant symmetries with the underlying QCD and allow
the investigations in a simplified framework. One example of such
effective models is the chiral quark-meson model which allows to
explore the chiral phase transition, see e.g.~\cite{Schaefer:2006sr}.

\section{Chiral effective quark-meson models}

Based on a previous analysis within an effective quark-meson model
with two quark flavors \cite{Schaefer:2007ep, Schaefer:2006ds,
  Schaefer:2004en} an extension to three quark flavors is done
straightforwardly which enables the investigation of the chiral
$SU(3)\times SU(3)$ symmetry restoration with temperature and quark
chemical potential including the axial $U(1)_A$ anomaly. Here we
briefly summarize some results of Ref.~\cite{Schaefer:2008hk} where
the three-flavor quark-meson model has been treated in mean-field
approximation. In this approximation the quantum and thermal meson
fluctuations of the grand potential are neglected while the
quarks/antiquarks are retained as quantum fields. The resulting
integration over the Grassmann fields yields finally the $T$- and
$\mu$-dependent quark/antiquark contribution $\Omega_{\bar q q}
(T,\mu)$ of the grand potential wherein the ultraviolet divergent
vacuum contribution has been neglected. After all the total grand
potential is a sum of $\Omega_{\bar q q} (T,\mu)$ and a meson
potential $U(\sigma_x, \sigma_y)$,
\begin{equation}
\label{eq:qmpot}
  \Omega(T,\mu) = \Omega_{\bar q q} (T,\mu) +U(\sigma_x, \sigma_y)\ ,
\end{equation}
where $\sigma_x$ and $\sigma_y$ denote the nonstrange and strange
condensates, respectively. The condensates are the corresponding
chiral order parameters and depend on $T$ and $\mu$. Note, that the
quark contribution also depends on these condensates implicitly via
the quark masses. Since we consider symmetric quark matter a uniform
quark chemical potential has been introduced.

\begin{figure}[tb]
\begin{minipage}[t]{0.3\textwidth}
\epsfig{file=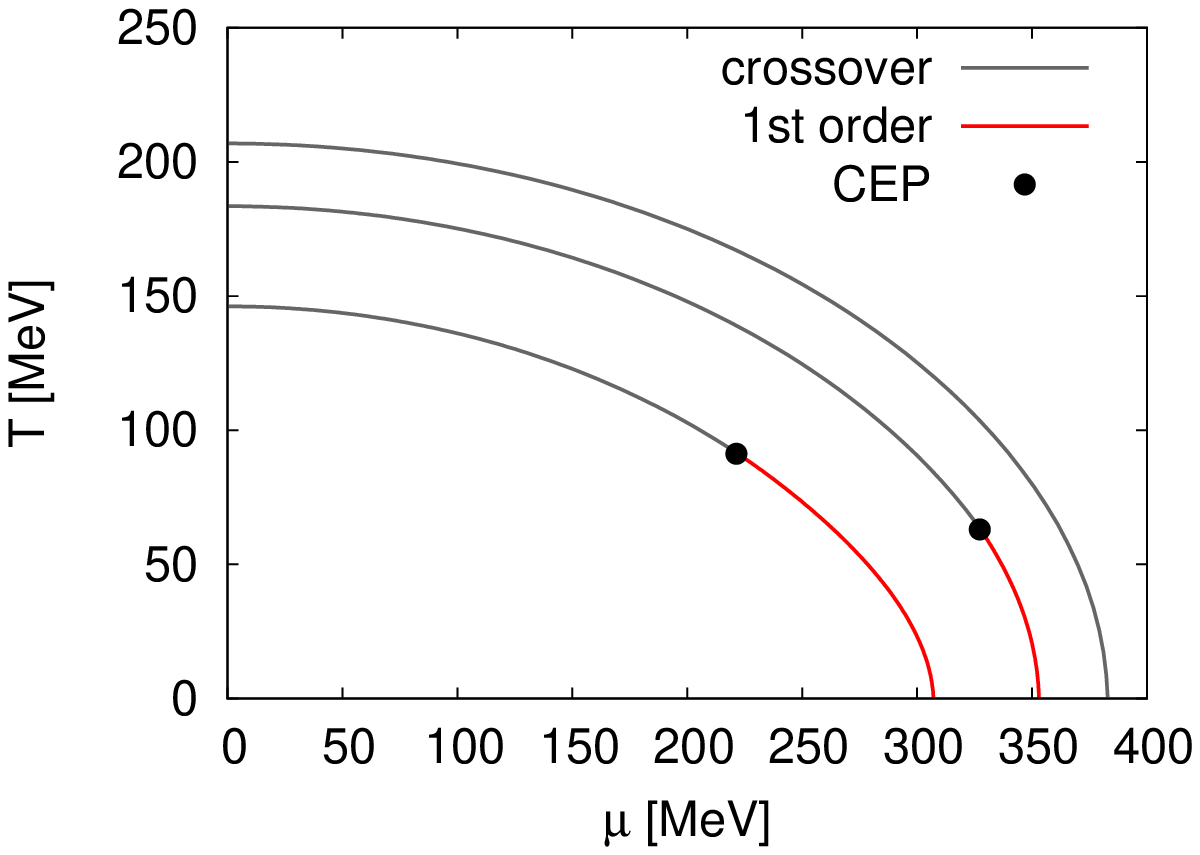,height=6cm}
\end{minipage}
\hspace*{3cm}
\begin{minipage}[t]{0.3\textwidth}
\epsfig{file=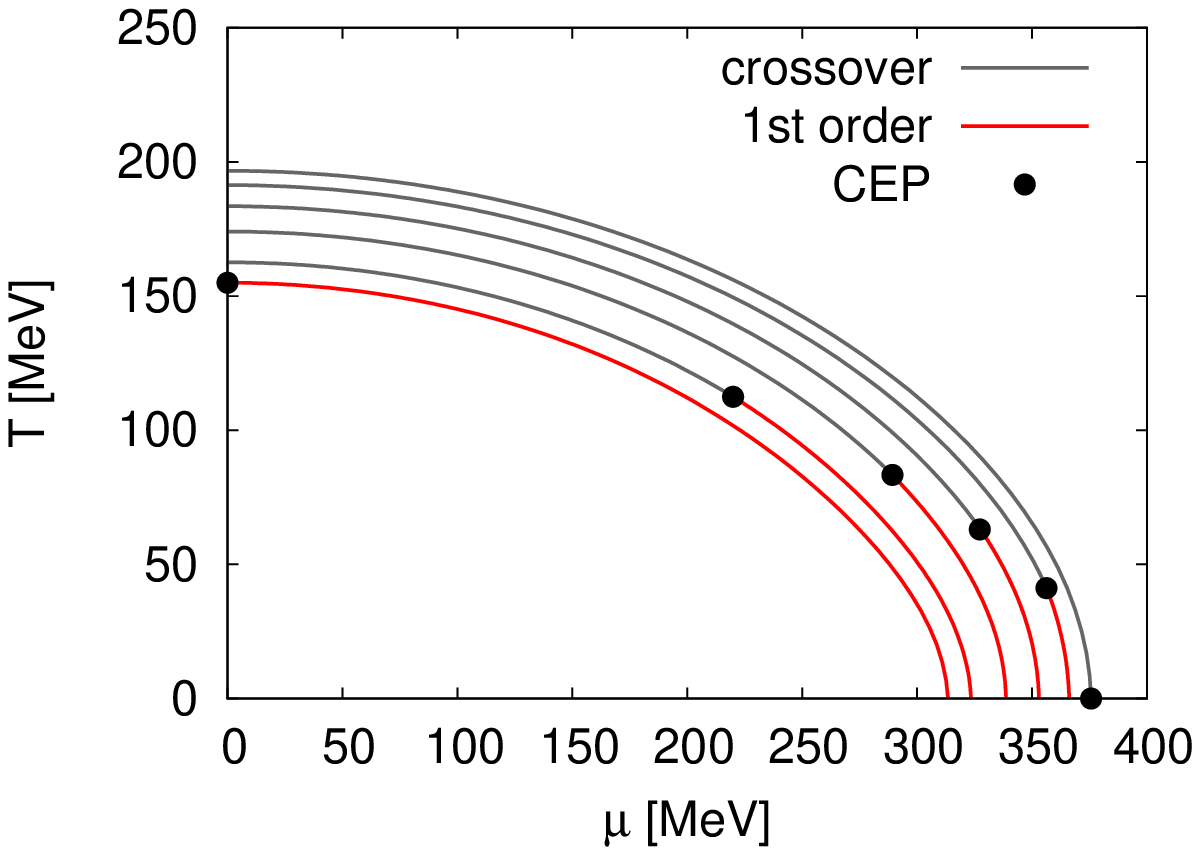,height=6cm}
\end{minipage}
\begin{center} 
\begin{minipage}[t]{\textwidth}
  \caption{ Left: Phase diagrams with axial anomaly for different
    values of the sigma mass: $m_{\sigma} = 600$ MeV (lower lines),
    $800$ MeV and $900$ MeV (upper line). Right: Phase diagrams with
    axial anomaly for $m_\sigma =800$ MeV but different pion masses
    $m_{\pi}/m_{\pi}^{*}= 0.49$ (lower line), $0.6, 0.8, 1.0, 1.2,
    1.36$ (upper line) where the ratio
    $m_{\pi}/m_{K}=m_{\pi}^{*}/m_{K}^{*}$ is kept fixed with
    $m_{\pi}^{*}=138$ MeV and $m_K^{*}=496$ MeV. \label{fig2}}
\end{minipage}
\end{center}
\end{figure}

The resulting phase diagrams with explicit $U(1)_A$ symmetry breaking
for three different values of $m_\sigma$ are shown in the left panel
of Fig.~\ref{fig2}.
For certain values of $m_\sigma$ a CEP is found. Compared to recent
lattice simulations, this point is located at smaller temperatures and
larger chemical potentials. Anyhow, its exact location in the phase
diagram cannot be predicted by effective models. The mass of the
$\sigma$ meson is one model input parameter which is poorly known
experimentally. We therefore use different input values for $m_\sigma$
in order to study its mass dependence. One sees for increasing
$m_\sigma$ that the location of the CEP moves towards the $\mu$-axis.
Already for $m_\sigma=900$ MeV the CEP disappears and the chiral phase
transition is a smooth crossover over the entire phase diagram.
Without axial anomaly almost no difference of the phase boundary and
hence of the location of the CEP is seen.

In Ref.~\cite{Struber:2007bm} a gauged chiral $U(2)\times U(2)$
symmetric linear sigma model without quarks within the 2PI resummation
scheme has been considered. If the influence of the vector mesons are
neglected the opposite behavior of the chiral transition as a function
of $m_\sigma$ is observed: at $\mu=0$ a crossover is found for a small
$\sigma$ mass and a first-order transition for a large $\sigma$ mass.
On the other hand, if vector mesons are incorporated the transition
leads to a more rapid crossover and brings one closer to the
second-order critical point.

Independent of the $U(1)_A$ symmetry breaking a first-order phase
transition in the chiral limit is expected. This behavior is shown in
the right panel of Fig.~\ref{fig2} where the phase diagrams including
the anomaly for varying pion and kaon masses are shown for
$m_\sigma=800$ MeV. For this figure a path in the $(m_\pi, m_K)$-plane
through the physical mass point towards the chiral limit has been
chosen by varying the pion mass while keeping the ratio $m_\pi/m_K$
fixed. On the one hand, for a pion mass $1.36$ times larger than the
physical one, the CEP lies exactly on the $\mu$-axis (for $m_\sigma =
800$ MeV) and the chiral transition is a smooth crossover over the
entire phase diagram. On the other hand, for decreasing pion masses
the location of the CEP moves towards the $T$-axis and for a pion mass
below half of the physical one the chiral transition turns into a
first-order one for all densities and no CEP exists any longer.

\begin{figure}[tb]
\begin{minipage}[t]{0.3\textwidth}
\epsfig{file=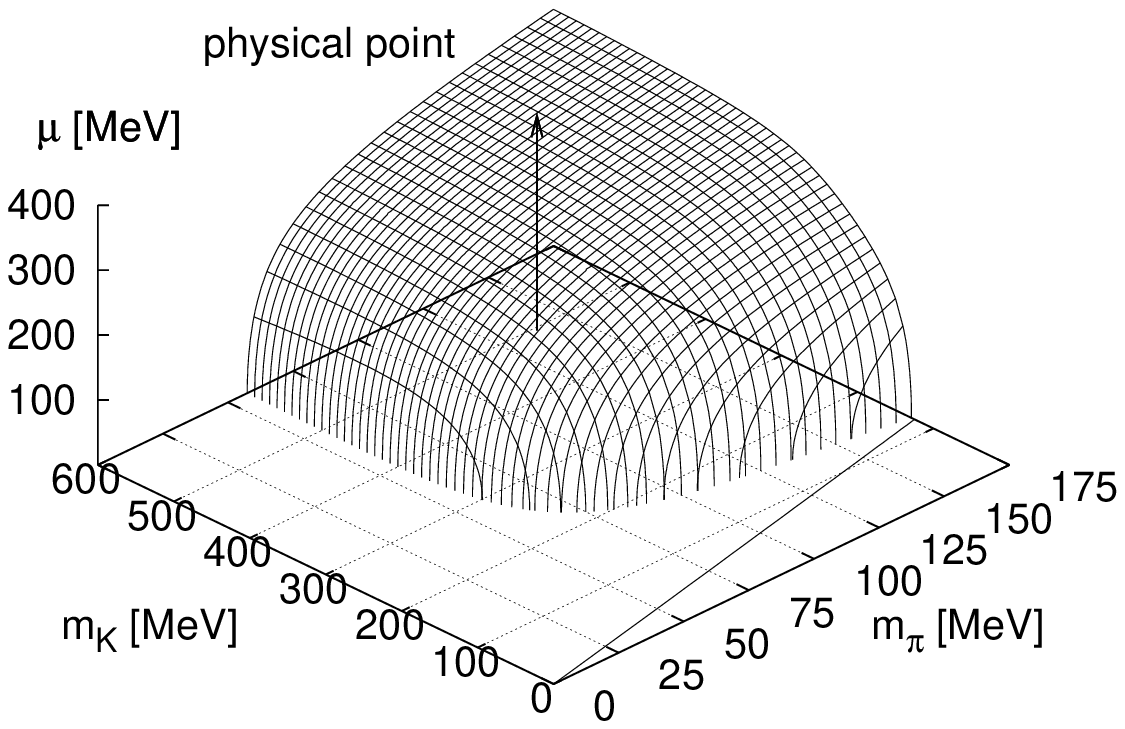,height=7cm}
\end{minipage}
\hspace*{3cm}
\begin{minipage}[t]{0.3\textwidth}
\epsfig{file=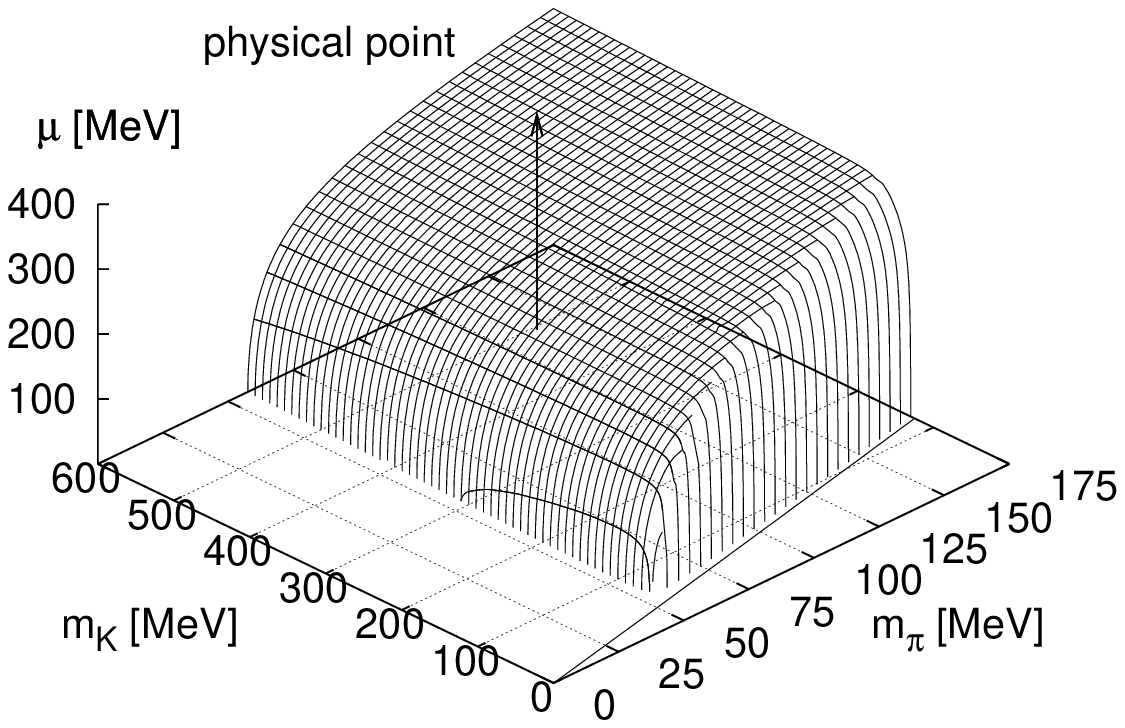,height=7cm}
\end{minipage}
\begin{center} 
\begin{minipage}[t]{\textwidth}
  \caption{Chiral critical surface as a function of the pion and kaon
    masses for $m_\sigma = 800$ MeV with (left) and without axial
    anomaly (right). The arrow points to the critical quark chemical
    potential at realistic pion and meson masses and is denoted as
    physical point.\label{fig3}}
\end{minipage}
\end{center}
\end{figure}

This behavior of the CEP excludes the nonstandard scenario described
in the introduction. The chiral critical surface which is defined by
the value of the critical chemical potential of the CEP for a given
mass pair ($m_\pi, m_K$), is evaluated in Fig.~\ref{fig3} as a
function of the pion and the kaon masses with (left) and without
(right) $U(1)_A$ symmetry breaking. For values of the chemical
potential above the surface the chiral transition is of first-order
while for values below the surface the transition lies in the
crossover region. With or without anomaly the surface grows out
perpendicular from the mass plane at $\mu=0$ and the tangent plane to
the critical surface has a decreasing slope for larger masses.
Consequently, the first-order region grows for increasing chemical
potentials. Since the critical chemical potential cannot grow
arbitrarily the surface must have a boundary at larger pion and kaon
masses which is not shown in the figure.

Furthermore, the effect of the $U(1)_A$ anomaly on the shape of the
surface is rather marginal for kaon masses greater than $400$ MeV.
This is reasonable since for larger kaon masses the strange sector
decouples effectively from the light nonstrange sector and the chiral
transition is basically driven by the light nonstrange particles. For
kaon masses smaller than $400$ MeV a considerable influence of the
anomaly on the shape of the critical surface is seen: without anomaly
the region of first-order phase transitions at $\mu=0$ is considerably
reduced.

\section{Emulating the Polyakov-loop dynamics}

So far only the chiral phase transition has been considered. Recently,
the quark-meson model could be combined with the Polyakov loop which
allows to investigate both, the chiral and the deconfinement phase
transition. In general, the Polyakov loop $\Phi$ is a complex scalar
field and serves as an order parameter for the
confinement/deconfinement transition in the quenched limit. Since it
is related to the free energy of a static test quark, it vanishes in
the confined phase where the free energy of a single quark diverges
and takes a finite value in the deconfined phase. It is linked to the
$Z(N_c)$ center symmetry of the $SU(N_c)$ gauge group. Thus, the
confining phase is center symmetric, whereas the center symmetry is
spontaneously broken in the deconfined phase.

\begin{figure}[tb]
\begin{minipage}[b]{0.3\textwidth}
\epsfig{file=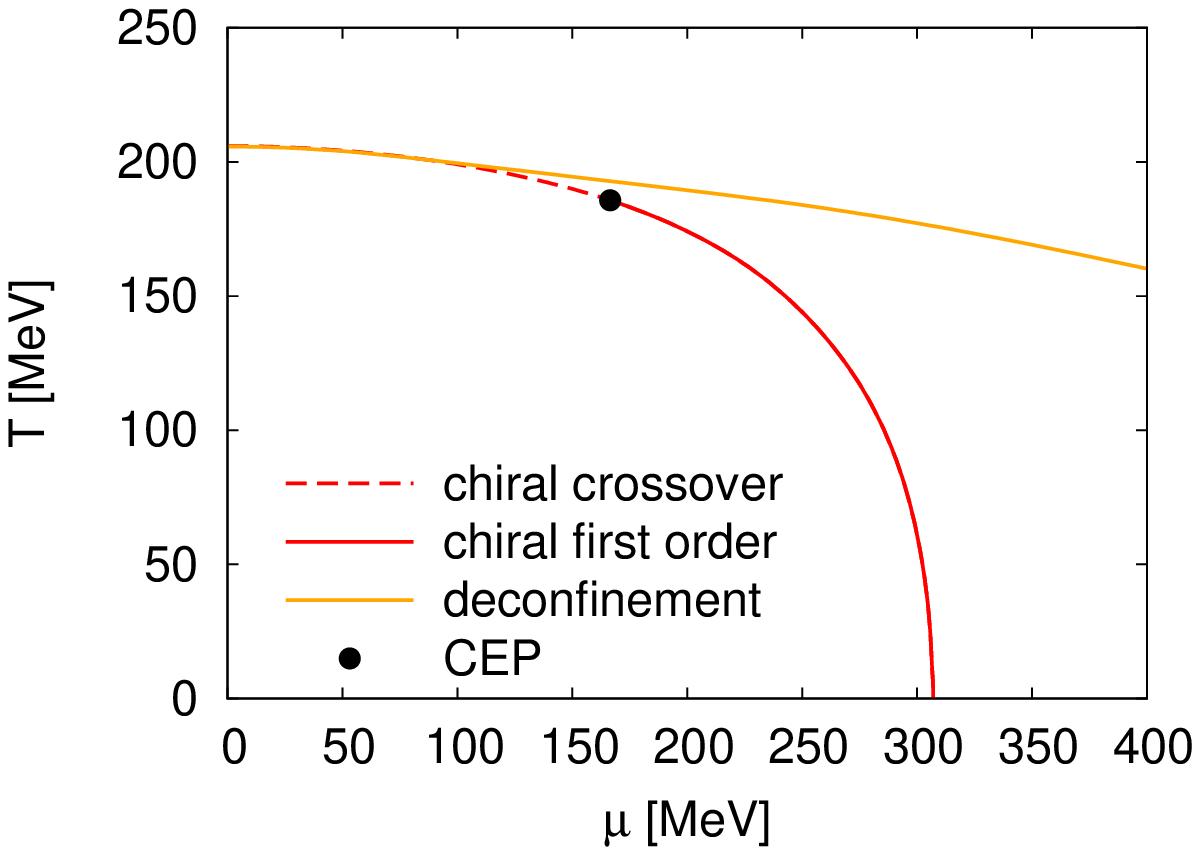,totalheight=6cm}
\vspace*{0.0mm}%
\end{minipage}%
\hspace*{3cm}%
\begin{minipage}[b]{0.3\textwidth}
\epsfig{file=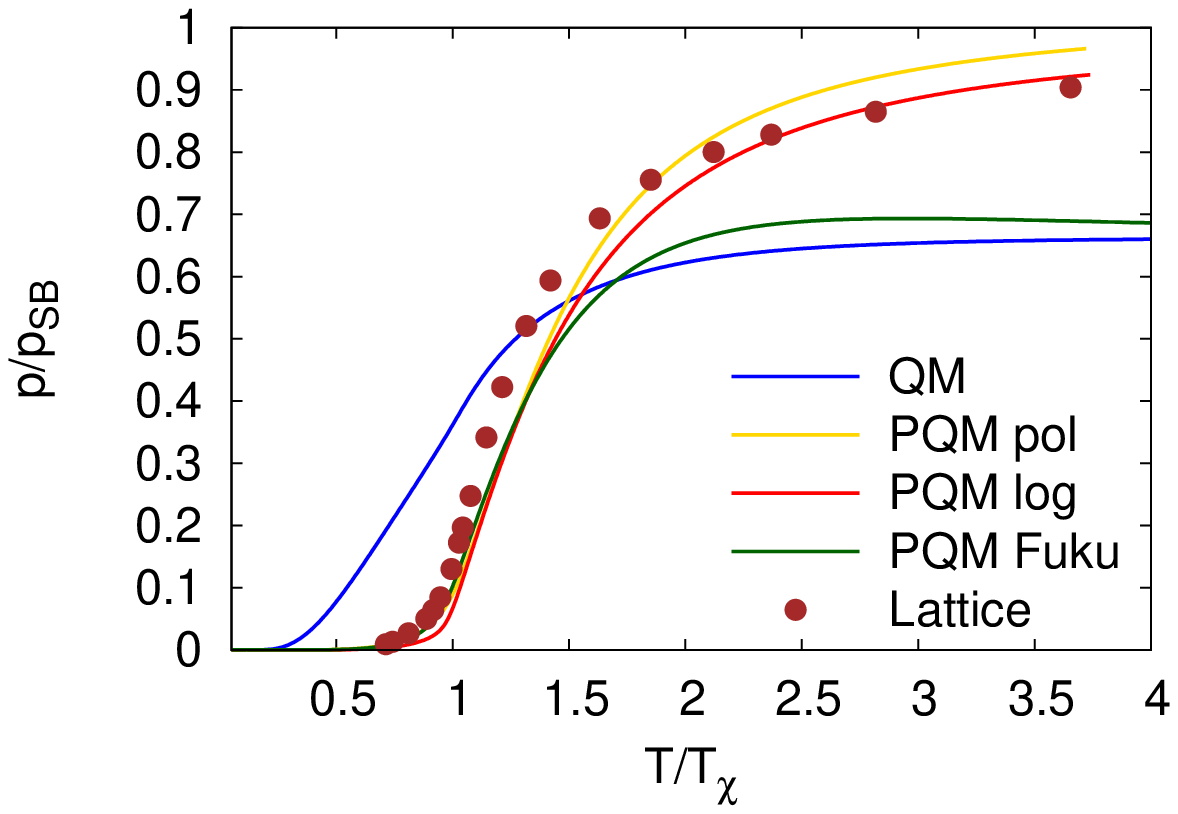,clip=,totalheight=6.5cm}
\end{minipage}
\begin{center} 
\begin{minipage}[t]{\textwidth}
  \caption{\label{fig:4}Left: Phase diagram for the three flavor PQM
    model with a logarithmic Polyakov loop potential. Right: Scaled
    pressure for the PQM model for three different Polyakov loop
    potentials (labeled as pol, log and Fuku) in comparison with
    recent lattice data for $2+1$ flavors \cite{Cheng:2007jq}. The
    scaled pressure without the Polyakov loop dynamics (QM) is also
    shown. }
\end{minipage}
\end{center}
\end{figure}

In the presence of dynamical quarks and nonvanishing chemical
potential it is not clear whether the Polyakov loop still serves as an
order parameter. In this case, the free energy does not diverge
anymore and the order parameter is always nonzero. Because the free
energies of quarks and antiquarks are different in the medium, $\Phi$,
related to quarks, and the Hermitian (charge) conjugate $\bar\Phi$,
related to antiquarks, will differ.

In pure Yang-Mills theory the real mean values $\Phi$ and $\bar\Phi$
are given by the minima of an effective Polyakov loop potential
$\mathcal U (\Phi,\bar\Phi)$ which can be constructed from lattice
data. Finally, the dynamical quark sector of QCD is included by
coupling the Polyakov loop to the quark sector of the quark-meson
model which leads to the Polyakov-quark-meson (PQM) model with an
interaction potential between quarks, mesons and the Polyakov loop
variables. For two quark flavors the PQM model has been introduced in
~\cite{Schaefer:2007pw}, wherein a polynomial ansatz for the Polyakov
loop potential has been used. Here, we extend the two flavor PQM model
to three quark flavors together with three different realizations of
the Polyakov loop potential $\mathcal U (\Phi,\bar\Phi)$
\cite{Fukushima:2008wg, Schaefer:2007pw, Roessner:2006xn}.

For three quark flavors the total grand potential of the PQM model is
a sum of three contributions
\begin{equation}
  \Omega(T,\mu) = \Omega_{\bar{q}{q}}(T,\mu;\Phi,\bar{\Phi}) + 
  U(\sigma_x,\sigma_y) + \mathcal{U}(T;\Phi,\bar{\Phi})\ ,
\end{equation}
where the quark/antiquark contribution $\Omega_{\bar{q}{q}}$ is
modified by the Polyakov loop variables. The mesonic contribution
$U(\sigma_x,\sigma_y)$ is the same as for the quark-meson model,
Eq.~(\ref{eq:qmpot}).

In Fig.~\ref{fig:4} (left panel) the resulting phase diagram of the
PQM model with three quark flavors for a logarithmic Polyakov loop
potential, adopted from~\cite{Roessner:2006xn}, is shown. The phase
boundaries are extracted from the peak in the temperature derivative
of the corresponding light nonstrange condensate and the Polyakov loop
variable. The model parameters are adjusted in such a way that both
phase transitions, the chiral and the deconfinement transition,
coincide at $\mu=0$ as is indicated by recent lattice
simulations~\cite{Cheng:2006qk}. In general, the chiral transition
temperature is shifted to higher values if the Polyakov loop is
included and a more rapid crossover is observed.

For increasing chemical potential both transition still coincide
initially but then start to deviate before the chiral critical point
is reached. The chiral transition always occurs below the
deconfinement transition. For small temperatures the chiral transition
becomes of first-order and both transitions are well separated. This
separation might be related to the existence of a quarkyonic phase but
work in this direction is still in progress.

The coupling of the quark dynamics to the Polyakov loop improves the
equation of state in the chirally broken phase at low temperatures and
densities. This is demonstrated in Fig.~\ref{fig:4} (right panel)
where the scaled pressure, normalized to the Stefan-Boltzmann
pressure, is shown at $\mu=0$ for three different realizations of the
Polyakov loop potential. In comparison, recent $2+1$-flavor lattice
data for $N_\tau = 6$~\cite{Cheng:2007jq} are also included and are in
reasonable agreement with our results. The suppression of the quark
contribution in the confined phase is clearly visible compared to the
pure QM model calculation.

\subsubsection*{Acknowledgments}
BJS is grateful to the organizers of the 30th International School of
Nuclear Physics in Erice, Italy, for the invitation and acknowledges
the European Physical Society (EPS) for the scholarship. MW was
supported by the BMBF grant 06DA123.


\end{document}